
\hskip 4.5in SNUTP 92-28
\title{\bf $\theta$ Bag and Quark Confinement}
\vskip 0.2in

\centerline{Jihn E. Kim}

\centerline{\sl Center   for   Theoretical   Physics    and
Department of Physics}
\centerline{\sl Seoul National University, Seoul 151-742, Korea}

\abstract{We propose $\theta$ bag through the wall
separating $\theta=0$ and $\theta=\pi$.  $\theta$ may  or
may not be a dynamical
field generating the wall.  For a massive pseudoscalar  $\theta$,
we present a two Higgs doublet model.   We  also
present  an  idea for quark confinement within this $\theta$  bag
scheme.}
\endpage

\def\s{{$SU(3)_c\times SU(2)\times U(1)$}}
\def\t{{$\theta$}}
\def\L{{$\Lambda_{QCD}$}}

Quarks appear to be confined in hadrons.  It is believed that the
nonabelian gauge group $SU(3)_c$ is the confining force of the
quarks.  At present a field theoretical explanation of
confinement is lacking even though the asymptotic freedom (or
infrared slavery) is suggested as an underlying mechanism [1].

A bag model description of hadrons through quarks confined
in a bag of radius $r_b$ has been suggested with some
success [2].  The chiral bag model modifies the original bag
idea including the imposed chiral symmetry [3].
These bag models, even though useful, are basically
phenomenological descriptions of hadrons,
and it is called for to have a
better realization of the bag from a fundamental Lagrangian.

In  this Letter, we suggest a confinement mechanism and try to
fill the gap between a field theoretic Lagrangian and bag models.
For confinement, we present an idea with a  plausible argument.
Because of the difficulty confronting the confinement  problem,
any new idea toward the solution is welcome.  An immediate
consequence of this
suggestion is a direct application to bag models.  Independently
from the confinement   mechanism, we also  present  a   field
theoretic realization  of axionic bag.  Even though the
confinement may not be realized as envisioned in the latter part
of this Letter, the ground states of quarks can be constructed as
color  singlet hadrons inside the \t\ bag in our field
theoretic model.  In the model the bag arises
from the domain wall separating two different QCD vacuum angles,
\t=0 and \t=$\pi$ where $\theta\equiv \theta_{QCD}+
\theta_{QFD}$.  $\lq$Domain wall' is not a good word for this
since these two vacua are not degenerate.  A better word is
$\lq$cliff' around an Arizona plateau,  but we will stick to
the word $\lq$wall'  in  this Letter because of its familiarity.
An  unbroken  nonabelian  group  necessarily
introduces a vacuum  angle \t\ due to instanton solutions [4].
Interactions   due   to  instantons  have  a  factor   $\exp (-8
\pi^2/g(\mu)^2)$. Therefore, instanton interactions are important
at low energy in asymptotically free theories.  If the theory  is
not  asymptotically  free,  the instanton interactions are not so
important for low energy physics.  This observation invites a
speculation for a possible connection between confinement and
instantons (or vacuum angle).  The \t\ dependence
of free energy gives a chiral symmetry breaking scale of order
\L.  In Fig. 1, the \t\ dependence is schematically presented.

Suppose  a  toy model in which $u$ and $d$ quarks  obtain  only
the current quark masses at \t=$\pi$ vacuum and obtain in addition
the constituent quark masses at \t=0 vacuum.  In this case there
can  exist a stable   hadron  due  to  the   energy
consideration.   For simplicity of the discussion, let us
neglect the wall energy, the constituent quark mass at \t=0, and
sea quarks and gluons inside a nucleon (\t=$\pi$).
Almost massless current quarks and gluons are supposed to be
confined in a \t=$\pi$ ball of radius $r_b$. (See Fig.2.)
$r_b$ is the bag radius which can
be determined by the minimum of the total energy of the ball.  If
the  sum of (three) constituent quark masses (for the case  of  a
baryon) at \t=0 vacuum  is  greater
than the sum of energies of the ball and current quark masses  at
\t=$\pi$ vacuum, formation of bag is energetically favorable.
The radius of \t=$\pi$ vacuum cannot be arbitrarily small due to
the uncertainty principle, viz. the kinetic energy of the
massless quark confined inside a ball of radius $r$ is $1/r$.
Thus, if the excess energy density of \t=$\pi$  vacuum relative
to that of \t=0 vacuum is $\rho$, the total energy of the ball is
$$E\ \simeq\ {4\pi\over 3}\rho r^3+{3\over r}\eqno (1) $$
where we counted 3 valence quarks inside a nucleon.  The  minimum
energy  condition, which must be smaller than  three  constituent
quark masses, determines the bag radius $r_b$
$$r_b\ \simeq\ \left({3\over 4\pi}\right)^{1/4}{1\over
\rho^{1/4}}\eqno (2)$$
For a reliable determination of the bag size, one should solve
strong  interaction  problem.  Here, we present an idea  how  the
hadron bag can be made from a field theoretic Lagrangian.  Within
this scheme, we also present an idea about the quark confinement.

\t\  can  be  understood  as a coupling  in a theory
without a massless quark or an axion.   In this case, the free
energy  description interpolates two different  theories  through
the bag.  This is called \t\ bag.  Along the surface of \t\ bag,
\t\  changes smoothly.  The merit of \t\ bag is that there is  no
contribution  to the surface energy density from the gradient  of
\t\ along the normal direction of the bag.

On the other hand, in a theory with a pseudoscalar particle $A$
coupling to the gluons through $AF^a_{\mu\nu}\tilde F^{a\mu\nu}$,
{\it different \t's merely mean different vacua}.
The field $A$ need not be an axion, but it can be called  axionic
bag.   The constraint of axionic bag is that it has a significant
surface energy contribution from the gradient of \t, $\sigma
\simeq  8m_\pi  f_\pi  F_A$  where  $F_A$  is  the  axion  decay
constant [5].   Thus, these axion models [5] cannot account  for
the axionic  bag, but a kind of variant axions [6] with a small
$F_A$ can be accomodated.

However, we present a model in which the $A$ field is
not necessarily an axion and furthermore the quark  masses inside
and outside the bag are different as in chiral bag models [3].
In a fully satisfactory theory of this type, this situation
should be proven from QCD dynamics identifying $A$ as a composite
pseudoscalar field.  Here, we take a rather modest  view  that
Higgs fields are responsible for the different quark
masses at different \t\ vacua.  In addition, my objective here is
not to solve the strong CP problem  and  do  not
introduce  the  Peccei-Quinn  symmetry.   The axion solution of
the strong CP problem has to be reinvestigated. Consider two
Higgs fields $H\ (Y=-\half)$ and $\phi\ (Y=\half)$.  The two
Higgs system has real symmetric quartic coupling matrices
($a_{ij}$  and  $b_{ij}$),  hermitian  quartic  coupling   matrix
($c_{ij}$), two real masses ($\mu^2_i$), and a complex mass ($
\mu_\epsilon^2$).  Without $c_{ij}$ and $\mu^2_\epsilon$ in the
potential, the potential has the familiar Peccei-Quinn
symmetry.[7]    With  the  Peccei-Quinn  symmetry,  the   vacuum
expectation value is not a function of \t.   Since a \t\ dependent
quark  mass  is needed in our scheme, we introduce  a
soft Peccei-Quinn symmetry violating $\mu^2_\epsilon$  term,  and
simplify  the  rest of couplings.
Thus, a toy model potential is taken as,
$$\eqalign{V\ =\ \lambda (H^\dagger H)^2&-\mu^2H^\dagger H
+\lambda_\phi (\phi^\dagger\phi)^2
-\mu_\phi^2\phi^\dagger\phi\cr
&-\mu^2_\epsilon\Big[\epsilon_{ij}H^i\phi^j+ c.c.\Big]}
\eqno (3) $$
There exist two $U(1)$ transformations : $H\rightarrow
e^{iy}H, \phi\rightarrow e^{-iy}\phi$, and $\{H,\phi\}
\rightarrow  e^{i\alpha}\{H,\phi\}$.  The first is a  part  of
the  $U(1)_Y$  gauge  transformation.  The  second contains an
anomalous (Peccei-Quinn) transformation and
the  needed  one.   Certainly, $V$ is not  invariant  under  this
anomalous transformation, and the phase field corresponding to
this transformation obtains a mass because of the $\mu^2_\epsilon$
term.  Let us assign vacuum expectation values of  the  form,
$\langle H^0\rangle =(v/\sqrt{2})e^{i\alpha}$ and
$\langle\phi^0\rangle=(u/  \sqrt{2})e^{i\alpha}$.  We can choose
$v$ and $u$ to be real and positive.  Then vacuum expectation
value of $V$  takes the form
$$\langle    V\rangle\    =\    {\lambda\over 4}v^4-{\mu^2\over
2}v^2+{\lambda_\phi\over 4}u^4-{\mu^2_\phi\over 2}u^2 -
\mu^2_\epsilon vu\cos (2\alpha)\eqno (4) $$
We  can interpret $2\alpha$ as \t\ if $v=u$.\foot{Otherwise,
\t\ is  a  bit more complicated.}
Because there is  no
Peccei-Quinn symmetry, \t=0 can be the minimum only by a fine
tuning of $\theta_{QCD}$.    Namely we fine-tuned $\theta_{QCD}$
such that the positions of minima of Fig. 1 and our electroweak
model coincide.   For the simple case
$\lambda=\lambda_\phi$ and $\mu^2=\mu^2_\phi$ which gives $v=u$,
one obtains
$v=0$ for \t=$\pi$ and $v=\sqrt{(\mu^2+\mu^2_\epsilon)/\lambda}$
for \t=0 in the parameter space $\mu^2_\epsilon>\mu^2$.
The  \t=$\pi$ vacuum has higher energy than \t=0  vacuum  by  an
amount
$${(\mu^2+\mu^2_\epsilon)^2\over 2\lambda}, $$
realizing  our constant positive energy density $\rho$
inside  the bag.\foot{Of course, the QCD vacuum energy  shown  in
Fig. 1 should be added to this.}
Quark masses are obtained from Yukawa couplings to $H$ and
$\phi$.    Thus  quarks are massless inside the bag  and  massive
outside the bag. An electroweak scale value $v\simeq  175$  GeV
and $\rho^{1/4}=O(100\ {\rm MeV})$ give
$\sqrt{\mu^2+\mu^2_\epsilon}\sim 80$ keV which cannot be
phenomenologically acceptable.  Furthermore, $v$ is too large  to
accomodate the bag surface energy of hadronic scale.
However, for this simple case we have shown a correct
direction of  symmetry  breaking.

Models   can    be
constructed  such  that  the Higgs masses can  be  raised  to  an
acceptable  value,  and  surface  energy  density  falls  in  the
hadronic energy scale.  For example, one may require that quarks
obtain  only  current quark masses inside the bag and  obtain  in
addtion the constituent quark mass outside the bag.  Suppose that
the  Yukawa couplings of $H$ give current quark masses and  those
of $\phi$ give constituent quark masses.  In this case, we may
require $\mu^2_\phi\ll \mu^2$ and $u\ll v$.  There exist such
solutions for parameters satisfying $\mu^2_\epsilon\mu
/\sqrt{\lambda}>2\mu^3_\phi /3\sqrt{3\lambda_\phi}$
so that $u\ne 0$ at \t=0 vacuum and $u=0$ at \t=$\pi$
vacuum and $v\sim 250$ GeV.  In this case,
$$V[\theta=\pi]-V[\theta=0]\ >\ 0$$
is also satisfied.
If one realizes the axionic bag of the above type
realizing an effective field $\phi$ in terms of quark bilinears,
one is closer to a simpler field theoretic
realization of axionic bag.  It is close to the phenomenological
chiral bag model.

In the \t\ bag (or the axionic bag) scheme, pion is better to be
interpreted as the pseudo-Goldstone boson arising from the
chiral  symmetry breaking by additional condition of $\langle
\bar qq\rangle\ne0$ in the \t=0 vacuum [8],  rather  than
as $q$ and $\bar q$ being put inside a bag.

Let  us proceed to discuss our idea of quark confinement  in  the
\t\  and axionic bag schemes.
The essential point of the $\lq\lq$confinement  problem" is how
to hide gluons inside the bag.\foot{One idea is that gluons are
massive outside the bag  while massless inside the bag [9].
We do not follow this line.}
Consider the action of gluons in a \t\ vacuum,
$$I\ =\ \int {\rm d}^4x\left(-{1\over 4}F^a_{\mu\nu} F^{a
\mu\nu}+{g^2\theta\over 32\pi^2}F^a_{\mu\nu}
\tilde F^{a\mu\nu}+g\bar q{\lambda^a\over 2}\gamma^\mu A^a_\mu q
+\cdots\right) \eqno (5)$$
where $a=1,2,\cdots,8$, $g$ is the \s\ gauge coupling constant,
and we have kept the quark-gluon coupling also.
{}From  the action, one usually
computes the energy of the classical gluon field.  The classical
gluon field is generated around a static isolated source (a quark
or a gluon).  (For massless quarks and gluons, one may consider
classical  field generated by a moving source.)  The estimate  of
the energy density of these classical fields is naively
anticipated to be $(1/2)({\bf E}_a^2+{\bf B}_a^2)$ at \t=0 vacuum.
The  \t\  term,  however, has to be  treated  more  carefully  to
exhibit the $2\pi$ periodicity of \t.

The QCD vacuum without quarks allows a free energy of the form
$$V[\theta]\ \simeq\ \Lambda_{QCD}^4(1-\cos\theta)
\eqno (6)  $$
To  guess  the form (6), we first note that
Hamiltonian density naively calculated from Eq. (5) is
$${\cal H}\ =\  \half    ({\bf    E}^2_a+{\bf B}^2_a)+
{g^2\theta\over 8\pi^2}{\bf E}_a\cdot {\bf B}_a\eqno (7) $$
In the Hamiltonian formulation, we integrate with ${\bf E}_a$
and ${\bf A}_a$ with a gauge choice $A^0_a=0$.  Here we treat
${\bf E}_a$ and ${\bf A}_a$ as independent variables, and
hence Eq. (7) does not reproduce the $2\pi$ periodicity of
\t.    Thus we argue that the Hamiltonian in the \t\ vacuum
must  exhibit an explicit \t=$2\pi$ periodicity.  In  this  case,
the \t\ vacuum has reproduced the symmetry property correctly.
A naive use of Eq. (7) amounts to an incorrect treatment of the
\t\  vacuum  property.  Namely, ${\cal H}$ cannot  be  determined
purely from the local interaction alone, but the whole vacuum
structure has to be taken into account.  In this Letter,
we just try a QCD Hamiltonian consistent with the \t\ periodicity
as, because we lack any better method,
$$\half f(\cos\theta)\left( {\bf E}_a^2+{\bf B}_a^2\right)
+{g^2\sin\theta\over 8\pi^2}
{\bf E}_a\cdot {\bf B}_a$$
where   we  inserted  $\sin\theta$  to  realize  the   $2\pi$
periodicity  explicitly.   More complex choice can be  made,  but
this  form  is  the simplest and consistent with the  CP
transformation property of the original Lagrangian.
We also introduced an arbitrary function $f(\cos \theta
)$ for generality.  Integration with respect to ${\bf E}_a$ gives
an  effective free energy of the form,
$$V[\theta]\ \sim\ \half\langle {\bf B}_a^2\rangle
\left(f(\cos\theta)-{g^4\sin ^2\theta\over 64\pi^2
f(\cos\theta)}\right) $$
Note that QCD vacuum can  have  gluon
field condensation, presumably with a positive sign $\langle
F_{\mu\nu}^aF^{a\mu\nu}\rangle\sim    0.012   {\rm    GeV}^4\cdot
\pi/\alpha_c$ [10]
which allows the possibility
$\langle {\bf E}^2_a \rangle >\langle  {\bf B}^2_a\rangle >0$.
Since  VEV  of ${\bf E}^2_a$ is nonzero, it is reasonable  to
assume VEV of ${\bf B}^2_a$ is also nonzero.  Thus
we can choose the vacuum expectation value of ${\bf B}_a^2$ to
be a positive constant for any value of \t, but we will integrate
out the ${\bf E}_a$ field.  Then it is
close to Eq.(6), but not
quite.   It has a periodicity  of $\pi$ due to
the $\sin^2\theta$ term.     The  $f$
function must be chosen so that the periodicity becomes $2\pi$.
For simplicity, we consider the QCD vacuum without quarks.
There are two simple choices : (a) $f=1-\cos\theta$ and
(b) $f=1+\cos\theta$.  Case (a) does not give the minimum of
$V$ at \t=0 due to the choice $\langle {\bf B}^2_a\rangle >0$,
and also is not consistent with our bag scenario.  We could  have
chosen $\sin (\theta/2)$ instead of $\sin \theta $ in front of
${\bf E}_a\cdot {\bf B}_a$ with $f(\cos \theta)=1$,
but  this case does not lead to the minimum of $V$ at  \t=0  with
our choice $\langle {\bf B}\rangle _a^2>0$.
So  it seems that Case (b) is the simplest choice.
Case (b) gives
$$V[\theta]\ \simeq\ -\Lambda_{QCD}^4
\left([1-{g^4\over 64\pi^2}]+[1+{g^4\over 64\pi^2}]\cos
\theta\right)\eqno (8) $$
A constant can be added in Eq. (8) to get Eq.
(6).   Since the well-known form Eq. (6) is obtained from our \t\
periodicity ansatz,
let the effective QCD Hamiltonian discussed above be valid
in the presense of \t\ term,
$${\cal H}_{eff}\ =\ \half(1+\cos\theta)({\bf E}_a^2+{\bf B}_a^2)
+{g^2\sin\theta\over 8\pi^2}{\bf E}_a\cdot {\bf B}_a
\eqno (9)$$
In this case, note that the field energy is doubled at the \t=0
vacuum compared
to the naive estimate Eq. (7), and it vanishes at $\theta=\pi$
vacuum.  We interpret that this  is the basic reason that
{\it the \t=0 vacuum expells  the  gluon fields}.
If  a massless quark is present, the ${\bf E}_a\cdot  {\bf  B}_a$
term  is  absent  and  there is no  need  to  introduce a \t\
dependent $f$ function.
In  this  case  we discard \t\ dependent  terms  in  Eq.(9).   If
massive quarks are present, the vacuum energy is a bit more
complicated, but the $2\pi$ periodicity of \t\ must be manifest
in ${\cal H}$.
This shows that the gluon source (quarks or gluons)
generate  \t=$\pi$  vacuum in the immediate neighborhood  of  the
source, to minimize the classical gluon field energy
(mainly that of ${\bf E}^2_a$), realizing
the interior of the bag.\foot{For simplicity, I compare \t=0 and
\t=$\pi$ only.}  The bag interior,
\t=$\pi$ vacuum, has a higher vacuum energy density compared to
the outside, \t=0 vacuum.  Therefore, if a color source, e.g. $q$
is created, the color source prefers to generate
a  one-dimensional structure of \t=$\pi$ vacuum just to make  the
volume smaller (to minimize sum of \t=$\pi$  vacuum
energy and the gluon field energy) and squeeze the
classical color field inside this one dimensional structure,
the so-called color flux string (more accurately a cylinder).
Fig. 3 depicts this situation for creating mesons (Fig.3(a)) and
baryons (Fig.3(b)).  The minimum energy consideration is the well
known creation mechanism of hadrons.  Taking
into account the kinetic energy of quarks (the uncertainty
principle guides an order) and vacuum energy, creation of mesons
(or baryons) is energetically favored by attaching $\bar q$
(or $q$ and $q$) to the flux line(s) and creating another
$q$ (or another $\bar q$ and $\bar q$).

We  can compare the energies of a color source in two cases
: (i) one in which $q$ is present inside the \t=$\pi$
ball (See Fig. 2.) and the rest is the \t=0 vacuum but the color
field smears into  \t=0 vacuum, and (ii) the other in which
$q$  and $\bar q$ are present inside the \t=$\pi$ ball (See  Fig.
2.) and the color field does not
smear into  \t=0  vacuum.  Considering the volume energy density
$\rho$ and kinetic energies of quarks only, we obtain
$$E_{min}\ \simeq\ {4\over 3}(4\pi)^{1/4}(1+4\pi g^2)^{3/4}
\rho^{1/4}\eqno (10.a) $$
for Case (i), and
$$E_{meson}\ \simeq\ {8\over 3}(2\pi)^{1/4}\rho^{1/4}
\eqno (10.b) $$
for  Case  (ii).  Up to 100 GeV, it is known that  $\alpha_c$  is
greater than 0.1, and hence numerically these energies compare as
$E_{min}>112\rho^{1/4}$   and   $E_{meson}\sim   4.22\rho^{1/4}$.
Certainly,   the   meson   creation  (e.g.   $\rho$   meson)   is
energetically  favored compared to creation of a  colored  quark.
For pion creation the preference is more striking.  In reality, a
quark is striken out from a nucleon.  See Fig. 4.  In this  case,
the  fast  moving  quark  will  create  \t=$\pi$ {\it vacuum
only} in the immediate neighborhood of the striken quark and
hence ${\bf E}_a$ field cannot go out to the \t=0 vacuum because
the field
contribition to the energy goes like $1/r^4$ which is enormous
because  $r$  is small.  Thus the color flux is  dragged  to  the
nucleon  creating  the  flux tube and  meson  is  created.   This
phenomenon is a kind of hadronization.
The energy argument prefers to
expell  the  gluon  field from \t=0 vacuum.  Thus an effective
mechanism for quark confinement in the \t\ or axionic bag is
realized.  It is effective in the sense that  an isolated quark
with finite energy can exist as given in Eq. (10.a).

In  conclusion, we presented an argument for quark
confinement  to realize \t\ and axionic bag models
from the \t\ term contribution to the classical gluon field
energy.  The \t=$\pi$ vacuum is the interior of hadrons and
\t=0 vacuum is the real vacuum.  The wall separating hadrons from
the  vacuum  is the \t\ bag or axionic bag.

\vskip 1in
\centerline{\bf Acknowledgments}

This  research  is  supported  in  part  by  Korea  Science   and
Engineering Foundation.

\endpage
\centerline{\bf References}

\pointbegin
S. Weinberg, Phys. Rev. Lett. {\bf 31} (1973) 494.

\point A. Chodos, R. L. Jaffe, K. Johnson, C. B. Thorn and V. F.
Weisskopf, Phys. Rev. {\bf D9} (1974) 3471.

\point  A.  Chodos and C. B. Thorn, Phys. Rev.  {\bf  D9}  (1975)
2733; G. E. Brown and M. Rho, Phys. Lett. {\bf B82} (1979) 177.

\point C. G. Callan, R. F. Dashen and D. J. Gross, Phys. Lett.
{\bf B63} (1976) 334; R. Jackiw and C. Rebbi, Phys. Rev. Lett.
{\bf 37} (1976) 172.

\point
S. Weinberg, Phys. Rev. Lett. {\bf 40} (1978) 223; F. Wilczek,
Phys. Rev. Lett. {\bf 40} (1978) 279;
J. E. Kim, Phys. Rev. Lett. {\bf 43}, (1979) 103; M. W. Shifman,
A. I. Vainstein, and V. I. Zakharov, Nucl. Phys.
{\bf B166}, (1980) 4933; M. Dine, W. Fischler, and M. Srednicki,
Phys. Lett. {\bf 104B}, (1981) 199; A. P. Zhitniskii, Yad. Fiz.
{\bf 31}, (1980) 497 [ Sov. J. Nucl. Phys. {\bf 31}, (1980) 360].
For a review, see, J. E. Kim, Phys. Rep. {\bf 150} (1987) 1.

\point R. D. Peccei, T. T. Wu and T. Yanagita, Phys. Lett. B172
(1986) 435;  L. Krauss and
F. Wilczek, Phys. Lett. B173 (1986) 189.

\point R. D. Peccei and H. R. Quinn, Phys. Rev. Lett. 38
(1977) 1440.

\point See, for example,  H.  Georgi, {\it Weak Interactions  and
Modern  Particle Theory} (Benjamin and Cummings Publishing Co.,
1984).

\point  See,  for example, W-Y. P. Hwang, Phys.  Rev.  {\bf  D29}
(1984) 1465.

\point M. A. Shifman, A. I. Vainstein and V. I. Zakharov, Nucl.
Phys. {\bf B147} (1979) 385; 448.

\endpage

\centerline{\bf Figure Captions}
\vskip 1.5cm
\noindent Fig. 1. A schematic view of $V[\theta]$.

\noindent Fig. 2. A ball of hadronic vacuum \t=$\pi$.

\noindent Fig. 3. Formation of a cylinder of color flux from
the energy consideration. ($a$) configuration can create
a meson by attaching $\bar q$, and ($b$) configuration
can create a baryon by attaching $q$ and $q$ at both ends.

\noindent Fig. 4. A quark $q$ striken out from a nucleon.  To
the momentum direction, the bag surface is close to $q$ and $r$
is very small.

\bye